# Lattice thermal conductivity and elastic modulus of XN$_4$ (X=Be, Mg and Pt) 2D materials using machine learning interatomic potentials

K. Ghorbani[1], P. Mirchi[1,2], S. Arabha[1,3], and Ali Rajabpour[1,4] [*], Sebastian Volz[5,6*]

[1] Advanced Simulation and Computing Laboratory (ASCL), Imam Khomeini International University, Qazvin, Iran.

[2] Department of Mechanical Engineering, K. N. Toosi University of Technology, Tehran, Iran.

[3] Department of Mechanical Engineering, Lassonde School of Engineering, York University, Toronto M3J 1P3, Canada.

[4] School of Nano Science, Institute for Research in Fundamental Sciences (IPM), Tehran, Iran.

[5] Institute of Industrial Science, The University of Tokyo, Tokyo 153-8505, Japan.

[6] LIMMS, CNRS-IIS IRL 2820, The University of Tokyo, Tokyo 153-8505, Japan.

## Abstract

The newly synthesized BeN$_4$ monolayer has introduced a novel group of 2D materials called nitrogen-rich 2D materials. In the present study, the anisotropic mechanical and thermal properties of three members of this group, BeN$_4$, MgN$_4$, and PtN$_4$, are investigated. To this end, a machine learning-based interatomic potential (MLIP) is developed on the basis of the moment tensor potential (MTP) method and utilized in classical molecular dynamics (MD) simulation. Mechanical properties are calculated by extracting the stress-strain curve and thermal properties by non-equilibrium molecular dynamics (NEMD) method. Acquired results show the anisotropic elastic modulus and lattice thermal conductivity of these materials. Generally, elastic modulus and thermal conductivity in the armchair direction are higher than in the zigzag direction. Also, the elastic anisotropy is almost constant at every temperature for BeN$_4$ and MgN$_4$, while for PtN$_4$, this parameter is decreased by increasing the temperature. The findings of this research are not only evidence of the application of machine learning in MD simulations, but also provide information on the basic anisotropic mechanical and thermal properties of these newly discovered 2D nanomaterials.

[*] Corresponding authors: A. Rajabpour, Email: rajabpour@eng.ikiu.ac.ir   S. Volz, Email: volz@iis.u-tokyo.ac.jp

## 1- Introduction

By the fabrication of graphene as the first two-dimensional (2D) material in 2004 [1], developing novel materials of this kind and investigating their properties and behavior have attracted considerable attention from researchers in various fields of science, ranging from chemistry and physics to engineering. Today, a wide range of 2D materials (e.g. phosphorene [2], borophene [3] h-BN [4]–[7], black phosphorene [8], [9], and $MoS_2$ [10], [11]) with distinguished mechanical, thermal, electrical and optical properties has been introduced. In addition to the properties of these materials, their geometry also makes them as promising structures with potential application in nanoelectronics and optoelectronics.

Generally, experimental study is the main approach in investigating the behavior and properties of materials and structures, but when it comes to micro/nanoscales', due to their submicron dimensions, using this approach encounters remarkable restrictions. So, theoretical and computer simulation-based methods in this regard have been remarked rapidly, particularly, after recent notable advances in computational tools. Molecular Dynamics (MD) simulation and density functional theory (DFT) are of the methods in this regard that have been successfully implemented in modeling of various systems including vibration behavior [12]–[17], mechanical and thermal properties [18]–[27], and electronic properties [28]–[31]. DFT is a time-consuming method providing the highest accurate results, while MD simulation is less time-consuming but its results firmly rely on the interatomic potential functions. Machine-learning methods provide an opportunity to develop specific potential functions for each desired simulation leading to highly accurate results in the range of DFT results with less computational cost. Mortazavi et al. [32]– [36] conducted various research on investigating the mechanical and thermal properties of micro/nano-scale materials through developing machine-learning interatomic potentials (MLIPs). Recently, Arabha et al. [37] published a comprehensive review paper on the application of MLIPs in the calculation of lattice thermal conductivity. In this study, thermal conductivity of different 2D and 3D materials obtained from various approaches including experiment, DFT, MD, and MLIP has been compared and the significance of developing MLIP was indicated. Also, different machine-learning techniques in creating interatomic potentials and their characteristics were presented in this paper. Zuo et al. [38] in a comprehensive comparison study analyzed the efficiency and the computational costs of different MLIPs including MTP, SNAP, qSNAP, NNP, and GAP for Cu and Ni fcc metals, Li and Mo bcc metals, and Si and Ge semiconductors which are the representative of various material/structural properties. Accordingly, it can be said that all MLIPs in comparison to classical interatomic potentials provide the highest accuracy in estimating the energies/forces and properties. Botu et al. [39] presented a workflow and five main steps in generating MLIPs.

The interfacial thermal conductance across $C_3N$, $C_3B$, $C_2N$, $C_3N_4$, and $C_3N_4$ carbon-based 2D structures was also investigated [40]. Mortazavi et al. [41] proposed a new computational approach to compute a phonon dispersion relation and analyze the dynamical stability of nanomaterials. This approach is based on a machine-learning interatomic potentials and provides faster and more efficient results than the DFT simulations. The usefulness of this approach has been presented for a wide range of low-symmetry and porous 2D nanomaterials.

Arabha and Rajabpour [42] investigated the thermal conductivity and Young's modulus of nitrogenated holey graphene ($C_2N$) using MLIPs. The significant dependency of thermal conductivity on length was reported in their work. They have also evaluated the development of MLIPs for the point-defected $C_2N$ structures. They have shown, while there is uncertainty in the stability of point-defected $C_2N$ structures using Tersoff classical interatomic potential, that the MLIP is able to well calculate the properties of these structures.

Recently, the synthesis of $BeN_4$ [43] introduced a new group of 2D materials called nitrogen-rich 2D materials with $XN_4$ chemical formulation in which X is the representative of metallic elements [28]. Thereafter, Mortazavi et al. [28] analyzed the stability and the mechanical, thermal, electronic, and phononic properties of a wide range of these 2D materials including $BeN_4$, $MgN_4$, $IrN_4$, $RhN_4$, $NiN_4$, $CuN_4$, $AuN_4$, $PdN_4$, and $PtN_4$ by conducting the DFT calculations. They have reported that among the observed materials only $BeN_4$, $MgN_4$, $IrN_4$, $PtN_4$ and $RhN_4$ have the required dynamical and thermal stability. Also, it has been shown that the mechanical properties of $BeN_4$, $MgN_4$, $PtN_4$ and $RhN_4$ are sensitive to the armchair and zigzag directions so that the armchair ones have a higher elastic modulus and tensile strength than those in zigzag structures. Berdiyorov et al. [44] investigated the anisotropic electronic properties of $BeN_4$ and $MgN_4$ using the DFT and non-equilibrium Green's functional methods. They have indicated that in comparison to zigzag direction, the armchair direction provides remarkably larger electronic charge transport capability. Tong et al. [45], [46] determined the anisotropic phonon and electron thermal conductivity of $BeN_4$ at room temperature and at high and ambient pressures utilizing the combination of Boltzmann transport equation and first-principle calculations. Cheng et al. investigated the lattice thermal conductivity of layered Dirac semimetal $BeN_4$ using the Boltzmann transport equation and first-principles three-phonon calculations [47]. Wang et al. [48] used first-principles simulations to investigate the thermal properties of $BeN_4$ and $MgN_4$.

In the present study, a machine learning-based interatomic potential is developed and employed in MD simulations for investigating the mechanical properties and thermal conductivity of $BeN_4$,

MgN$_4$, and PtN$_4$ 2D monolayers shown in Figure 1. The main steps in this regard are as follows. Firstly, the ab-initio molecular dynamic trajectories are obtained as the subsample set, and then, the interatomic potential is trained over this subsample using the momentum tensor potential (MTP) machine learning method. Finally, utilizing the trained interatomic potential in the MD simulation, the desired mechanical and thermal properties are calculated.

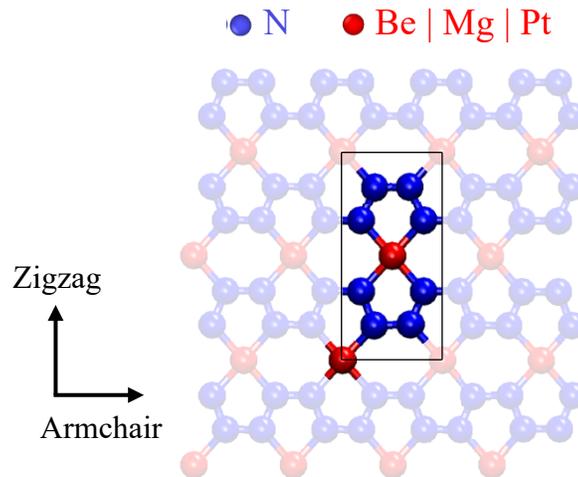

Figure 1. Atomic structure of BeN$_4$, MgN$_4$, and PtN$_4$ 2D monolayers and their corresponding armchair and zigzag directions.

2- **Methodology**

In this section, the process of extracting the machine learning-based interatomic potentials using the ab-initio molecular dynamic (AIMD) results and MTP method proposed by Shapeev [49], and utilizing them in classical MD simulations to calculate the mechanical properties and thermal conductivity of BeN$_4$, MgN$_4$, and PtN$_4$ will be discussed.

*2.1 Ab-initio simulation*

The aim of implementing ab-initio simulations is to produce the necessary dataset sources for training the MLIPs. To this end, structures are firstly optimized for seven different strains between 0-14% and nine various temperatures between 100-1000 K. It should be noted, all simulation was performed 1000 times steps of 1 fs, in every strain and temperature.

*2.2 Interatomic potential training*

Here, the MLIPs are developed using the MTP [49]. According to this method, the interatomic potentials can be defined as a multiplication of inertia tensors of radial polynomial functions. In order to train the parameters of MTP, the difference between the obtained results from the first principle calculations (i.e. energy, forces, and stresses) and the predicted results is minimized [49], [50]:

$$\sum_{m=1}^{M}\left[w_e(E_m^{AIMD} - E_m^{MTP})^2 + w_f \sum_{i}^{N}|f_{m,i}^{AIMD} - f_{m,i}^{MTP}|^2 + w_s \sum_{i,j}^{N}|\sigma_{m,ij}^{AIMD} - \sigma_{m,ij}^{MTP}|^2\right] \quad (1)$$
$$\rightarrow min$$

In this equation, $m$ is the number of training set configurations and $N$ is the total number of atoms. $E_m$, $f_{m,i}$, and $\sigma_{m,i}$ are respectively energy, force, and stress terms, and AIMD and MTP superscripts indicate the corresponding values obtained from AIMD modeling and MTP calculations. Also, $w_e$, $w_f$, and $w_s$ are the positive weighted coefficients showing the significance of energy, force, and stress expressions, and are considered to be 1, 0.1, and 0.001, respectively. For the initial training subsample sets, 10% of the AIMD calculated trajectories at first are considered in this regard. That is due to the fact that the correlation of the calculated trajectories are done in a short-range time. Then, MTP is trained over these initial subsample sets. By assessing the accuracy of the trained MTP over all of the AIMD trajectories, the subsample sets are updated by the high degree of extrapolation trajectories [51], and again MTP is trained over these new subsample sets. The algorithm of training the MLIPs is indicated in Figure 2. The ultimate trained MTP is considered as the MLIP and used in MD simulations.

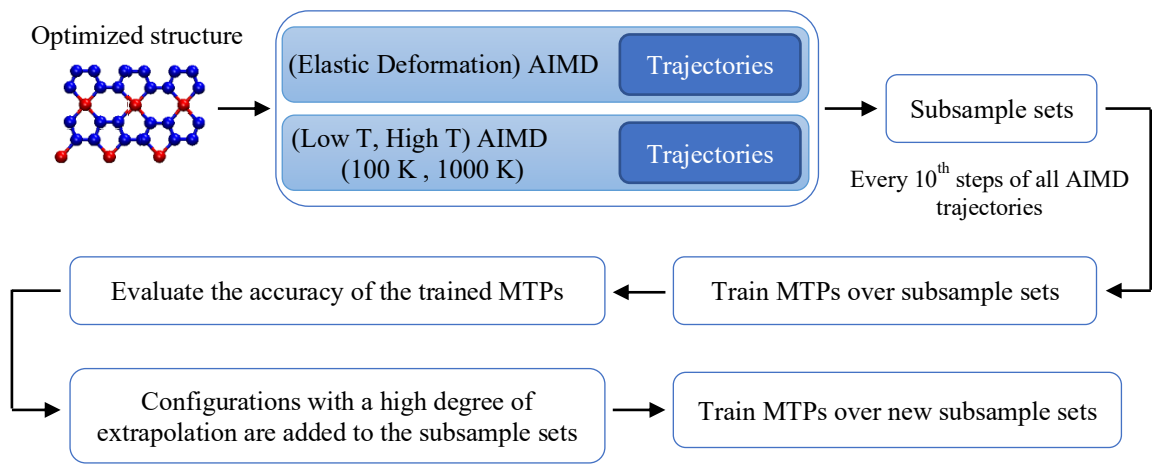

Figure 2. Machine-learning interatomic potentials (MLIPs) training algorithm. [42]

*2.3 Molecular dynamic simulations*

In this paper, MD simulations are performed using a Large-scale Atomic/Molecular Massively Parallel Simulator (LAMMPS) package [52], and the MTP trained potential as the interatomic potential, with 0.5 fs time-step and 300K average temperature. Also, the boundary conditions are periodic which remove the surface effects at boundaries and minimize the influence of finite-length.

The lattice thermal conductivity of desired materials is calculated based on the non-equilibrium molecular dynamics (NEMD) [53][54][55][56] method. To this end, at first by the use of Nose-Hoover barostat and thermostat (NPT) at room temperature, the nanostructures are relaxed for 2.5 ps. Then, the heat flux is imposed in the equilibrated structures by creating a temperature gradient between the defined hot and cold regions at the end sides of the structures, as shown in Figure 3.

The added and subtracted average energy from the two baths are calculated in the microcanonical ensemble (NVE) for 3ns. At the end, the thermal conductivity is calculated using Fourier's law of heat conduction as follows:

$$\kappa = -q'' / \frac{dT}{dx} \tag{2}$$

In which $q''$ is the heat flux through the structure and $\frac{dT}{dx}$ is the temperature gradient within it.

In order to compute the mechanical properties, the simulation box at first is stretched in the direction of loading at a specific engineering strain rate of $1 \times 10^9 \ S^{-1}$, and then by plotting the stress-strain curve, the slop in the linear region is reported as the Young's modulus of the structure. Based on the virial theorem, the stress values are obtained as follows [57]:

$$S = \frac{1}{V} \sum_{a \epsilon V} \left[ -m\vec{v}_a \otimes \vec{v}_a + \frac{1}{2} \sum_{a \neq b} (\vec{r}_{ab} \otimes \vec{F}_{ab}) \right] \tag{3}$$

in which, $S$ represents the stress tensor and $V$ is the volume of structure. $m$ and $\vec{v}_a$ are the mass and velocity vector, respectively. Also, $\vec{r}_{ab}$ and $\vec{F}_{ab}$ indicate the position and force vectors between atoms $a$ and $b$, and $\otimes$ denotes the outer product.

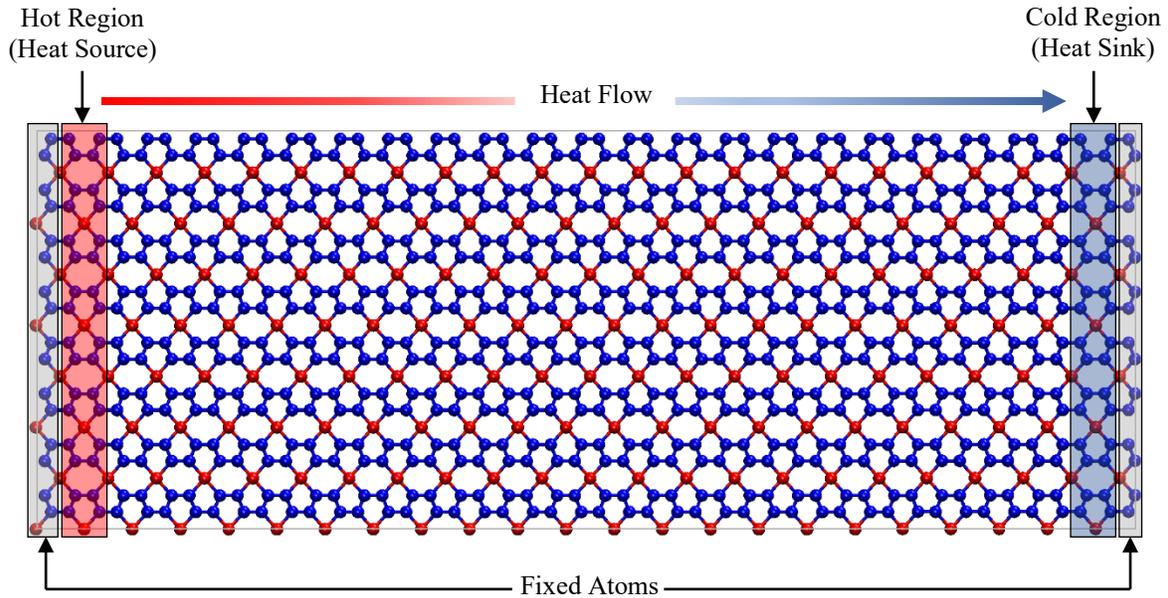

Figure 3. Schematic illustration of the NEMD setup for the calculation of thermal conductivity

### 3- Results and discussion

The main focus of the present research is the calculation of thermal and mechanical properties of $BeN_4$, $MgN_4$, and $PtN_4$ monolayers using the MD simulation and MLIP. The acceptable data set for training the MLIP must consist in all the possible trajectories that a system may experience in MD simulation. So, to generate training data sets, a wide range of temperature, from 100 K to 1000 K, is considered for the AIMD simulations. This range of temperature provides trajectories with both long-wavelength phonons, related to low-temperature conditions, and high-frequency optical modes, related to high-temperature conditions. Also, the high-temperature AIMD trajectories are needed for the consideration of large local deformations that probably happen in NEMD simulations of long-length structures.

Before presenting the thermal and mechanical properties, the dynamical stability of the considered structures is examined through the phonon dispersion relations (PDR) obtained from MLIP-based

MD simulations using mlip_phonopy scripts [41]. As illustrated in Figure 4, no imaginary frequencies appear in the PDR, which confirms the dynamical stability of $BeN_4$, $MgN_4$, and $PtN_4$ 2D monolayers. Moreover, PDR predicted by DFT is shown in Figure S1 (Supplementary data).

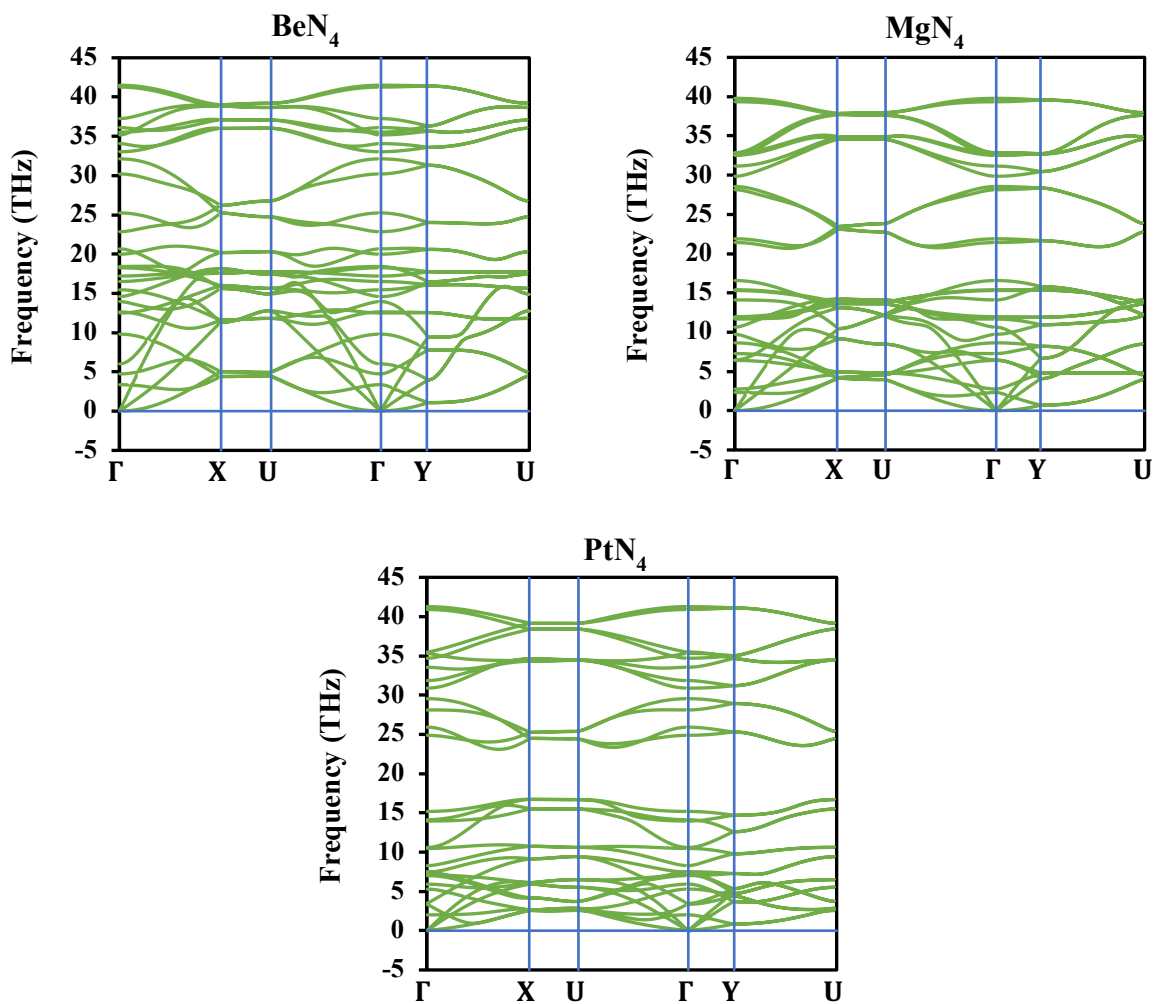

Figure 4. Phonon dispersion relations of $BeN_4$, $MgN_4$, and $PtN_4$ 2D monolayers.

*3.1 Thermal properties*

By implementing the trained MLIP in NEMD simulations, the lattice thermal conductivity of $BeN_4$, $MgN_4$, and $PtN_4$ 2D nano sheets in the armchair and zigzag directions has been investigated.

In Figure 5 the variation of thermal conductivity versus length in both armchair and zigzag directions is shown. As can be seen in this figure, for all these three materials and in both directions, the thermal conductivity at short lengths is length-dependent so that by increasing the length, the thermal conductivity increases. But, as the length increases, this dependency decreases and thermal conductivity converges to a specific value at longer lengths. The values of thermal conductivity in both directions are presented in the same figure.

From a comparative point of view, it can be seen that the BeN$_4$ monolayer has the highest thermal conductivity in both directions with ~140 and ~ 107 W/m-K in armchair and zigzag directions respectively, while the lowest values of this property in armchair and zigzag directions belong to the PtN$_4$ monolayers with ~ 115 and ~ 41 W/m-K respectively. Also, the change rate of thermal conductivity relative to length is different in armchair and zigzag directions and there is a significant difference in the values of thermal conductivity between these two directions. In other words, it can be stated that in thermal applications, the orientation of these 2D materials is of great importance. It should be noted that the thermal conductivity at infinite length, $k_\infty$, is approximated through fitting of the well-known equation $1/k_L = (1 + \Lambda/L)/k_\infty$ [58] on the calculated thermal conductivity at finite lengths, $k_L$. In this equation $\Lambda$ is the effective phonon mean free path.

The observed thermal behavior is justifiable by analyzing the phonon group velocity as the main criteria of lattice thermal conductivity variation. To this end, the phonon group velocity of BeN$_4$, MgN$_4$, and PtN$_4$ are plotted in Figure 6. Also, in this figure the phonon group velocity of graphene as a benchmark is presented. The phonon group velocities at low-frequency modes are generally considered as the main heat carrier. As it is obvious in this figure, at low-frequency modes, BeN$_4$ and PtN$_4$ respectively have the highest and lowest phonon group velocities that is the indicative of their thermal conductivity. This result is completely aligned with the obtained lattice thermal conductivity.

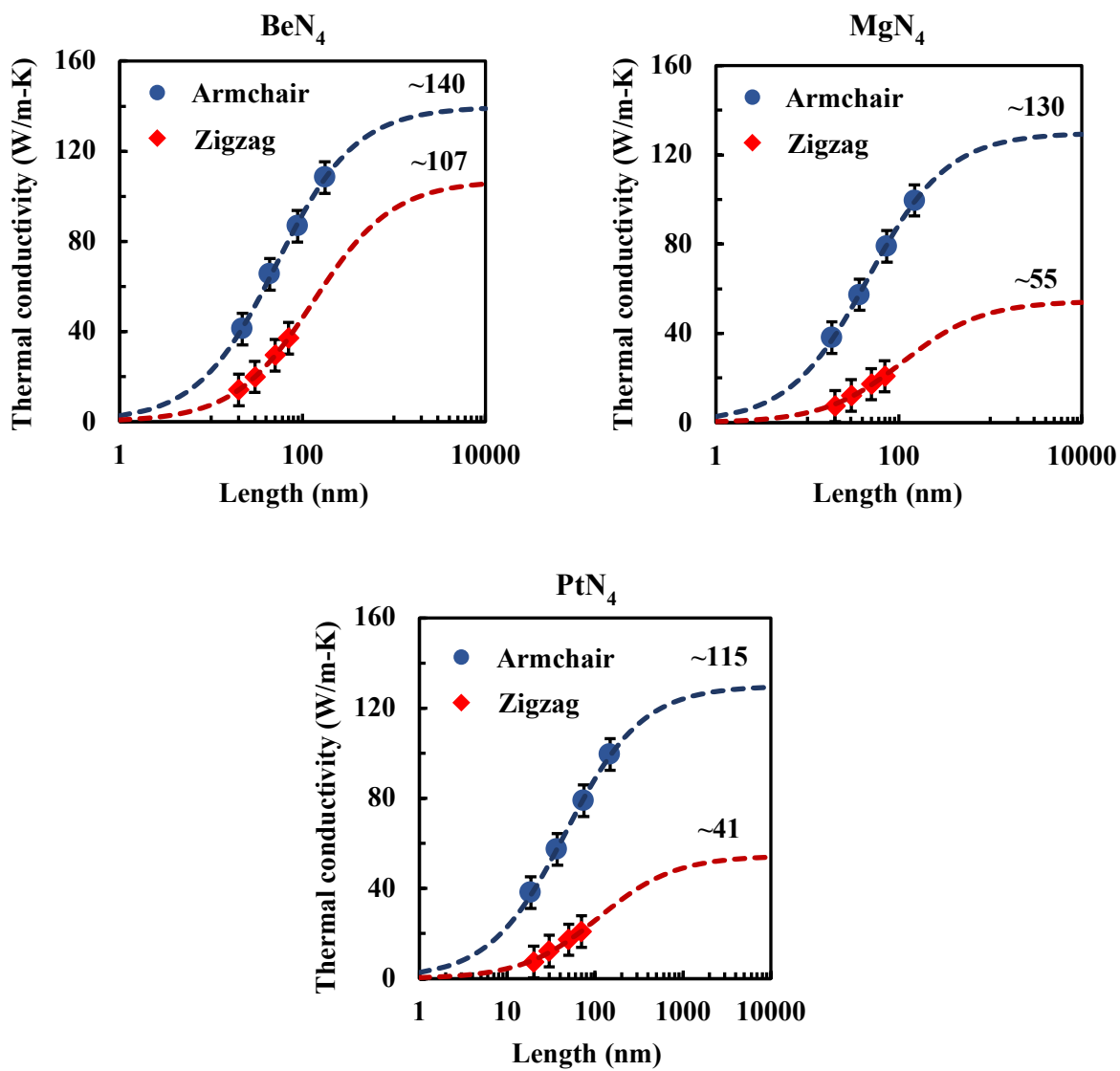

Figure 5. The variation of lattice thermal conductivity as a function of length at 300K.

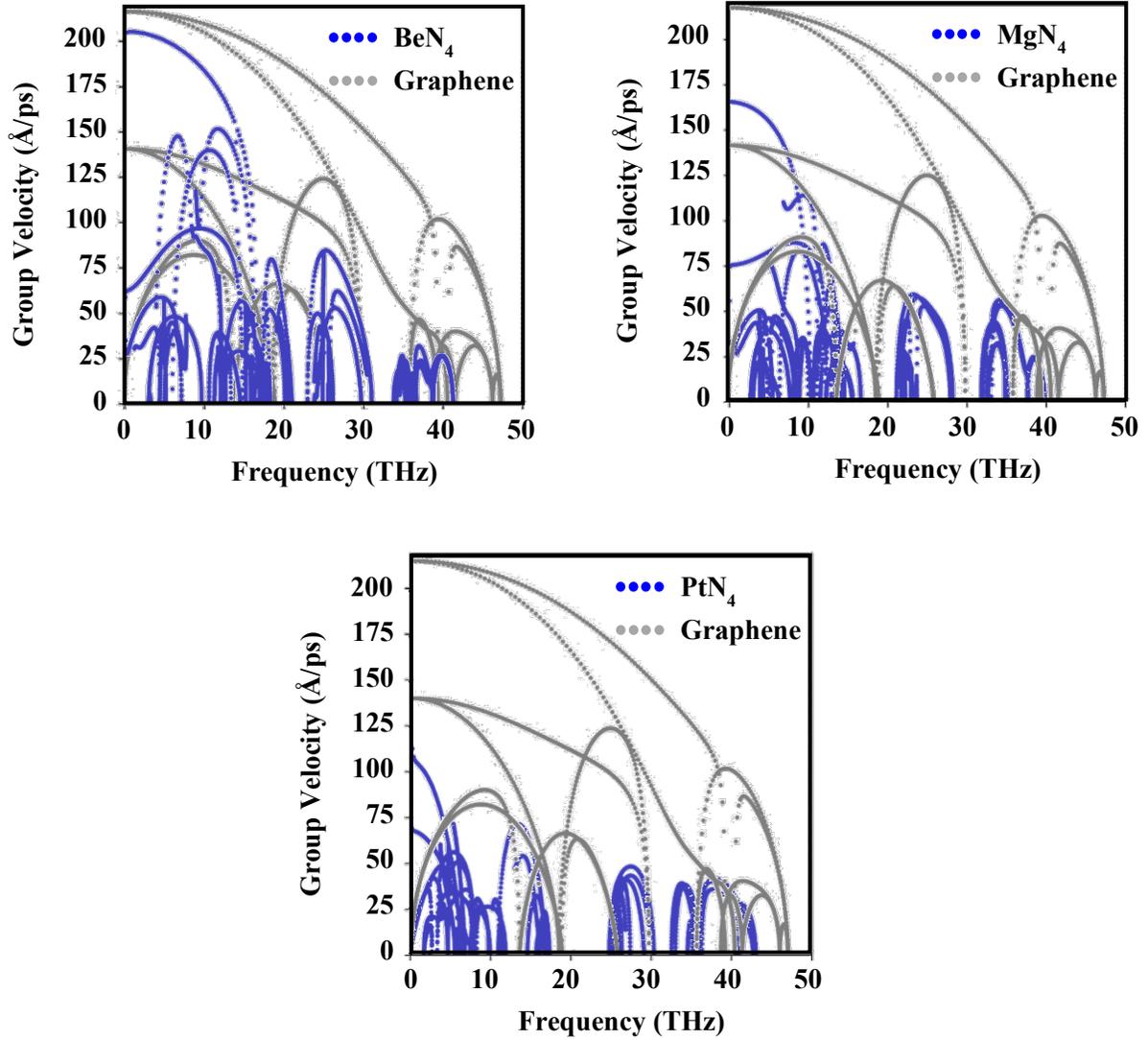

Figure 6. Phonon group velocity. Comparison of BeN$_4$, MgN$_4$, and PtN$_4$ with graphene.

*3.2 Mechanical properties*

In this subsection, temperature and chirality-dependent mechanical properties of BeN$_4$, MgN$_4$, and PtN$_4$ 2D materials obtained from classical MD simulations with the trained interatomic potential will be discussed. To this end, the uniaxial stress-strain relation in armchair and zigzag directions of these materials (see Figure 1) at room temperature is evaluated and shown in Figure 7. It should be noted that the thickness of structures is assumed to be 3.06 Å. According to this figure, it can be said that for these 2D materials, the ultimate strength in the zigzag direction is generally lower than those of armchair ones. Also, as can be concluded from this figure, the ultimate strength of BeN$_4$, MgN$_4$, and PtN$_4$ in the armchair direction are approximately close together (with about 70, 76, and 76 GPa respectively). The same conclusion is not valid in the zigzag direction so that just the ultimate strength of BeN$_4$ and MgN$_4$ ($\approx$ 40 GPa) are similar in this direction while for PtN$_4$ the value of this property is about 70 GPa. The reason for such a behavior and generally for the anisotropic mechanical behavior of these structures is rooted in the dominant bonds that have to be broken during the uniaxial loading. As it is clear in Figure 1 , the N-N bond is perfectly aligned in the armchair direction so in the case of uniaxial loading along this direction, it can be said that initiation of the bond failure as the criterion in calculating the ultimate strength depends mostly on N-N bonds regardless of the type of metallic atoms (Be, Mg, and Pt) in the structure. On the other hand, in the zigzag direction, the Be-N, Mg-N, and Pt-N bonds play a dominant role so that the type of metallic atoms has an influence on the ultimate strength.

In contrast to MgN$_4$, the fracture strain in the zigzag direction is higher than their armchair counterparts. Also, it can be deduced from this figure that MgN$_4$ and BeN$_4$ have the highest and lowest values of fracture strain respectively.

In Figure 8, the variation of elastic modulus versus temperature has been presented. The dependency of elastic modulus on chirality and temperature is obvious in this figure. As illustrated in this figure, armchair structures have higher values of elastic modulus than the ones of the zigzag structures, and by increasing the temperature, the elastic modulus decreases in both armchair and zigzag directions. Moreover, it can be noticed that both armchair and zigzag structures of MgN$_4$ have the lowest values of elastic modulus, while interestingly, the highest values of this property are associated with the BeN$_4$ in armchair direction and PtN$_4$ in the zigzag direction. The elastic modulus of BeN4, MgN4, and PtN4 along the armchair (zigzag) directions presented in Figure 9 corresponds to a 13% (23%), 11% (17%), and 13% (10%) difference from those reported in Ref [28], respectively. The different input data sizes and strain ranges employed in AIMD may be the cause of these discrepancies.

In order to examine the dependency of anisotropic behavior on temperature, the variation of anisotropy relative to temperature is plotted in Figure 10. Here, anisotropy is defined as $\frac{E_a - E_z}{E_a}$ where $E_a$ and $E_z$ are the elastic modulus along the armchair and zigzag directions respectively. As can be seen in this figure, for BeN$_4$ and MgN$_4$, the anisotropy is about 0.4 and almost constant at any temperature, while for PtN$_4$, the anisotropy is less than 0.2 and decreases by increasing the temperature. This behavior is also obvious in Figure 8 more specifically for PtN$_4$. So, it can be stated that BeN$_4$ and MgN$_4$ display an anisotropic mechanical behavior at each temperature but PtN$_4$ shows an isotropic mechanical behavior at high temperatures.

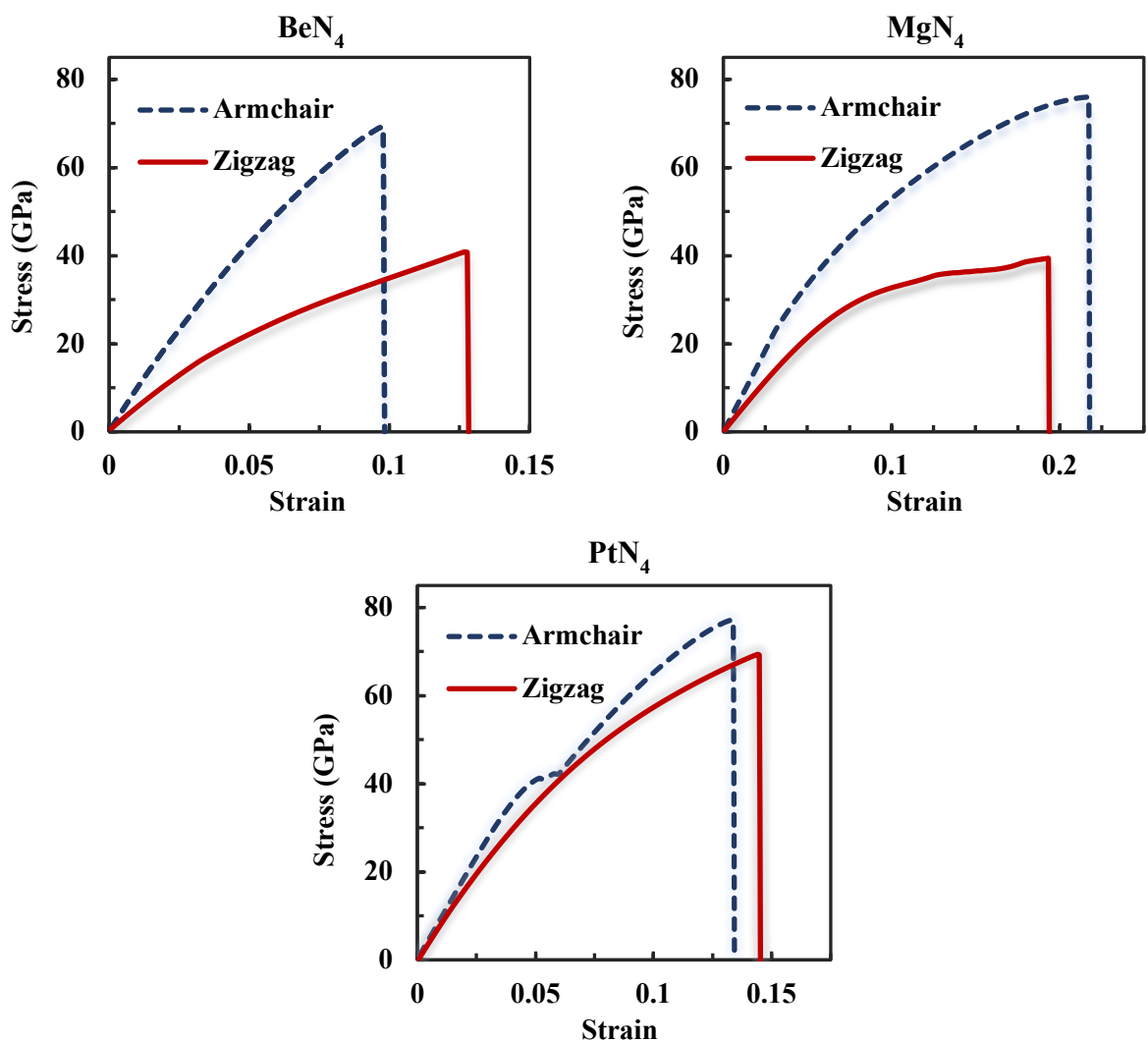

Figure 7. Stress-strain diagram under uniaxial tensile load at 300K.

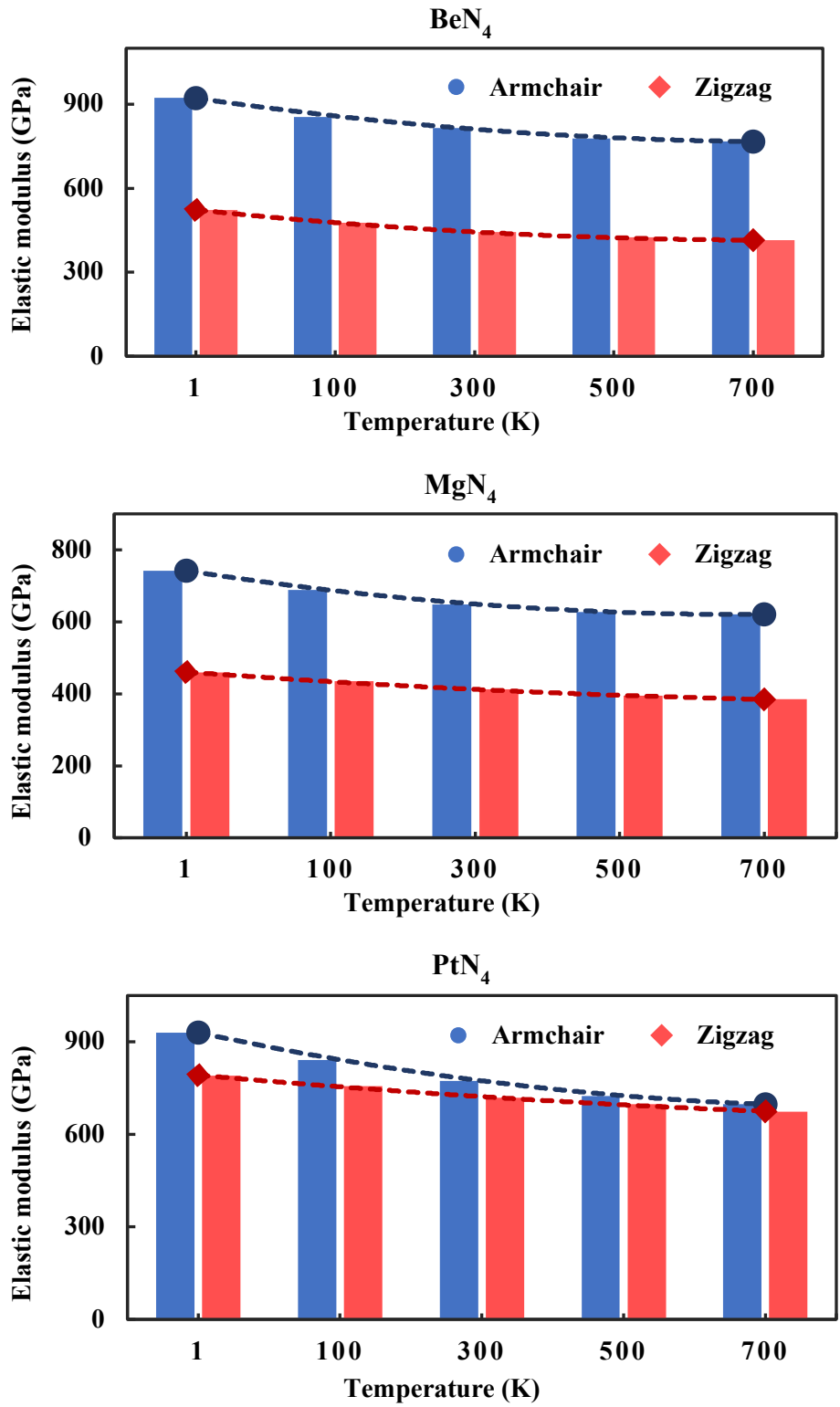

Figure 8. The variation of elastic modulus as a function of temperature.

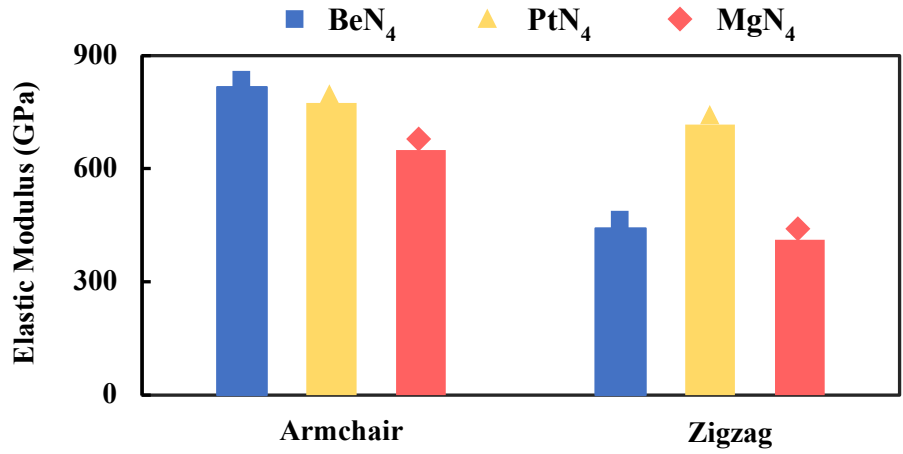

Figure 9. Comparison of elastic moduli of $BeN_4$, $MgN_4$, and $PtN_4$ in armchair and zigzag directions at 300K

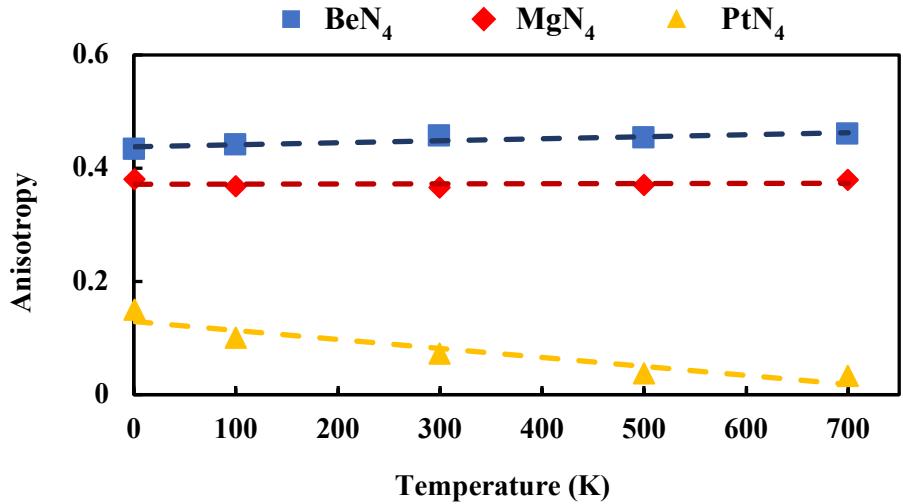

Figure 10. Dependency of elastic modulus anisotropy on temperature.

4- **Conclusion**

In summary, the present study provided details on the anisotropic mechanical and thermal properties of $BeN_4$, $MgN_4$, and $PtN_4$ monolayers, a new group of 2D materials. This aim was achieved by developing a MLIP using MTP method, and MD simulation. The anisotropic elastic modulus and lattice thermal conductivity are concluded from the results in such a way that armchair orientation leads to higher elastic modulus and thermal conductivity than the ones in the zigzag orientation. Also, the anisotropy ($\frac{E_a - E_z}{E_a}$) in BeN$_4$ and MgN$_4$ is independent of temperature and is approximately constant. However, increasing the temperature reduces anisotropy in PtN$_4$. Moreover, thermal conductivity increases by increasing the length and at long lengths, it reaches a constant value so that BeN$_4$ present the highest thermal conductivity in armchair and zigzag directions, and PtN$_4$ exhibits the lowest values in both directions. On the other hand, the order from highest to lowest values of elastic moduli follows BeN$_4$, PtN$_4$, and MgN$_4$, in the armchair direction and PtN$_4$, BeN$_4$, and MgN$_4$ in the zigzag direction, while this trend for thermal conductivity is BeN$_4$, MgN$_4$, and PtN$_4$ in both directions.

Furthermore, the present study is an evidence of the application of machine learning approach in developing interatomic potential functions and use them in classical MD simulations yielding the highest accuracy with less computational costs.

## Acknowledgment

This work is based upon research funded by Iran National Science Foundation (INSF) under project No. 4002089.

25